**Professor Ion DOBRE, PhD**
**The Bucharest University of Economic Studies**
**E-mail: dobrerio@yahoo.com**
**Ionuț JIANU, PhD Candidate**
**E-mail: ionutjianu91@yahoo.com**
**Associate Professor Dumitru Alexandru BODISLAV, PhD**
**E-mail: alex.bodislav@ase.ro**
**Associate Professor Carmen Valentina RĂDULESCU, PhD**
**E-mail: cv_radulescu@yahoo.com**
**Lecturer Sorin BURLACU, PhD**
**The Bucharest University of Economic Studies**
**E-mail: sburlacu@man.ase.ro**

# THE IMPLICATIONS OF INSTITUTIONAL SPECIFICITIES ON THE INCOME INEQUALITIES DRIVERS IN EUROPEAN UNION

*Abstract.This paper aims to review the different impacts of income inequality drivers on the Gini coefficient, depending on institutional specificities. In this context, we divided the European Union member states in two clusters (the cluster of member states with inclusive institutions / extractive institutions) using the institutional pillar as a clustering criterion. In both cases, we assesed the impact of income inequality drivers on Gini coefficient by using a fixed effects model in order to examine the role and importance of the institutions in the dynamics of income disparities.The models were estimated by applying the Panel Estimated Generalized Least Squares (EGLS) method, this being weighted by Cross-section weights option. The separate assessment of the income inequality reactivity to the change in its determinants according to the institutional criterion represents a new approach in this field of research and the results show that the impact of moderating income inequality strategies is limitedin the case of member states with extractive institutions.*

### 1. Introduction

Inequality still continues to represent a global issue in its various forms: payment, income, wealth, gender, opportunity. This paper focuses on the review of the income inequality in EU member states and on the examination of its drivers, since the economic and financial crisis has increased the social challenges for European citizens. In particular, this phenomenon has been covered from economic, social and institutional perspective. Nevertheless, empirical evidence proved to be more robust in terms of the implications of economic and social developments on the dynamics of income inequalities.

This paper aims to complete the economic literature in the field by setting up an analysis framework for the member states of the European Union to facilitate the comparison of the impact differences of income inequality drivers on this phenomenon, depending on the institutional specificities.

The main motivation for selecting this theme is based both on the actuality of this phenomenon and on the insufficient coverage of the institutions impact on the dynamics of the income inequality in the relevant published literature.

Moreover, from the institutional perspective, it is more relevant to capture the results of the institutions and not just the inputs. Also, the negative externalities of the capitalism on inequality phenomenom are one of the main reasons for which the economic system may fail on long run.

The objective of this research is to determine the implications of institutional specificities on the impact exercised by the drivers of income inequality, its achievement depending on the following specific objectives:

✓ clustering the European Union member states according to the institutional criteria published by the World Economic Forum in two groups of countries: member states with inclusive institutions and member states with extractive institutions;

✓ estimating the impact of the income inequality drivers on the Gini coefficient for each cluster defined previously;

✓ comparing the impact differences resulted from the models estimated.

This research is structured as follows:

1. second section: the main results of the relevant published literature in this field;
2. third section: the methodology used for achieving the research objective;
3. fourth section: results and interpretation;
4. fifth section: conclusions and recommendations.

Finally, this research will provide recommendations for policy intervention, depending on the differences between the impact of the income inequality drivers associated to each model.

## 2. Literature review

Research results in this area provide both robust and uncertain evidence, depending on the indicators used in the assessment. Regarding the evolution of the inequality phenomenon, the International Monetary Fund (2017) proved that in the last 30 years, 53% of countries experienced an increase in income inequality (catched by Gini coefficient) by at least two deviation points. This indicator is also the most relevant measure for the level of income inequality, its methodology being made available by Gini (1912).

Even if the recent published literature has provided evidence to facilitate the inclusion of other factors (globalisation, low immigration, automation /

technological progress) on the income inequality drivers list, some economists believe that theinstitutional arrangements influencing the distribution of economic growth benefits among the population, may cause changes in the dynamics of income gap (Acemoglu and Robinson, 2002; Stiglitz et al., 2009).

Chong and Gradstein (2007) estimated the impact of institutional quality on income inequality using a Panel window and found that improving the quality of institutions is associated with reducing income inequality between income and vice versa. Other authors have pointed out that inequalities do not arise as a result of the natural forces of the markets and are a consequence of the institutions through which those has been established (Clark and Kavanagh, 1996).

Kaplan and Raugh (2013) see increasing inequality as a phenomenon that depends on the marginal labour productivity, technological progress or the differences between non-market mechanisms that allow political and economic elites to extract excess rents, which damages the overall economic system through manipulation, corporate governance, social norms, respectively the tax and regulatory system. Moreover, the income gap is also determined by the political process, when the economic agents succeed to obtain favors from political decision-makers (subsidies, regulations or preferential laws).

However, the published literature is not broad enough and further evidence is needed to fully address the relationship between quality of institutions and income inequality, which strengthens the motivation of choosing this research area. This paper does not aim to follow the same methodology used by the authors willing to estimate the impact of the quality of institutions on income inequality. Its objective is to identify the specificities of the determinants of the income inequality, depending on the institutional characteristics. In order to cover the relevant theoretical framework, it is also essential to highlight the empirical evidence on the sign of the control variables coefficients.

Engerman and Sokolof (2002) argued that the main determinants of income inequality developed in colonial regimes are the historical factors. Niehues (2010) provided additional evidence to confirm the strong link between Gini's historical values and the current level of the indicator. Following the use of the Panel GMM method, with a robust specification of instruments, the author has demonstrated that the increase in the Gini lagged by one year leads to an increase in the current inequality with about 65% of the change, this being the highest impact exerted by the exogenous variables on its dynamics. Also, the impact of the autoregressive term has been found to be significant at a 1% significance threshold.

Theoretically, in an economy, extreme inequality may coexist with a minimum level of poverty and vice versa, but in practice this phenomenon is not frequent as these indicators are statistics calculated on the basis of the same household income distribution, which creates a mechanical link between them. Lynch et al. (2000) provided the reason for the high correlation between inequality and poverty, highlighting the fact that societies that prefer inequality are those who are not dedicated to poverty reduction. The authors pointed out that social norms

61

and values that make it possible to deepen economic inequality also facilitate the rise of poverty.

Dabla-Norris et al. (2015) indicated that the low quality of the educational system and the asymmetry of educational competencies are among the most important determinants of inequality in advanced countries. Ward et al. (2009) examined the relationship between unemployment and income inequality and found that a positive shock on unemployment rate leads to a deepen of income inequality.

Regarding the analysis of the relationship between social spending and income inequality, Niehus (2010) found that increasing social spending leads to a reduction in income disparities. This effect has been also confirmed by Fuest et al. (2010).

Some economists have also focused on studying the relationship between access to credit and income inequality, and have shown that high borrowing costs make it difficult for people at risk of poverty to access loans (Perotti, 1996). Thus, this instrument loses one of its main integrative function: facilitating the access of low-income people to education. The positive effect of private sector credit (% of GDP) on the dynamics of the Gini coefficient was also confirmed by Jaumotte et al. (2008).

## 3. Methodology

In this section, we described the methodology used to identify the role of institutions in the dynamics of income inequality recorded in the European Union member states. This paper starts from the hypothesis used by Acemoglu and Robinson (2012), according to which the quality of the institutions is a relevant determinant of the level of prosperity or poverty / inequality. The authors associate the extractive institutions with high poverty rates and the inclusive institutions with low poverty rates.

Firstly, we looked for the main indicators that can be used as a proxy for the quality of institutions. The identification of such an indicator may be methodologically sensitive, as the main statistical sources, organizations or institutions that publish such data are built on the surveys basis. Therefore, it should be pointed out that such indicators assess the target audience's perception on the purpose of the survey. In this context, we chose to use an indicator reflecting more the perception on the results of the institutions (outputs) and not the inputs, this approach having the role to catch better the efficiency of government in several areas. More specifically, we used the scores of the institutions pillar published by the World Economic Forum in the Global Competitiveness Report 2017-2018 for the European Union member states - the data from this publication are the values reported for 2016. The limits set by the organisation for this indicator are 1 (minimum score reflecting the lowest quality of institutions) and 7 (maximum score reflecting the highest quality of institutions).

In order to ensure the feasibility of the comparison between European Union member states in various institutional areas, we have used the sub-indices of the institutional pillar published by the World Economic Forum as follows (*Table 1*):

All sub-indices are numerically represented on a scale of 1-7, according to the main indicator of institutions quality, with the exception of the sub-indicator no. 21 which is numerically represented on a scale of 1-10 (where 1 reflects a minimum quality of the institution and 10 reflects a maximum quality of the institution).

**Table 1. Structure of the institutions pillar**

| Institutions pillar | | | |
|---|---|---|---|
| 1. Property rights | 7. Favoritism in decisions of government officials | 13. Business costs of terrorism | 19. Efficacy of corporate boards |
| 2. Intellectual property protection | 8. Efficiency of government spending | 14. Business costs of crime and violence | |
| 3. Diversion of public funds | 9. Burden of government regulation | 15. Organized crime | 20. Protection of minority shareholders interests |
| 4. Public trust in politicians | 10. Efficiency of legal framework in setting disputes | 16. Reliability of police services | |
| 5. Irregular payments and bribes | 11. Efficiency of legal framework in challenging regulations | 17. Ethical behaviour of firms | 21. Strength of investor protection |
| 6. Judicial independence | 12. Transparency of government policymaking | 18. Strength of auditing and reporting standards | |

*Source: Own processings using World Economic Forum data, The Global Competitiveness Report 2017-2018*

Estimating the impact of institutional quality on income inequality may be subjected to uncertainty given that the institutional pillar methodology reflects more the Institution's perceptions of the Forum. In this context, we preferred to use this indicator as a clustering criterion and not as a potential explanatory factor of income inequality, which facilitates the separation of member states in two clusters, depending on the national quality of institutions: the cluster of member states with inclusive institutions and the cluster of member states with extractive institutions. The categorisation of states in the clusters above-mentioned was made using the median of the institutions scores reported for each EU Member States. The use of the median provided the possibility to construct two clusters with an equal number of countries, which creates the premises for a more relevant comparison between these two groups of states. In this context, we included the member states with institutions scores above 4.29 in the cluster of member states with inclusive institutions, while the member states recording scores below the median were placed in the group of countries with extractive institutions. Although Brexit started to become a reality, this research focuses on the European Union with 28 member states for ensuring the similarity of observations between clusters.

Subsequently, we estimated the impact of the explanatory factors of income inequality on the Gini coefficient for each cluster, and finally we compared the impact differences between the estimated models to highlight the specificities of each institutional cluster. For this purpose, we used annual data for 2010-2016

63

period, given that the variables extracted from Eurostat and World Bank do not cover a large period of analysis for all countries analysed. Since 2016 corresponds to the clustering criterion used, this became the upper limit of the time period analysed. The structure of the clusters and data required the use of panel data for both group of countries, each model including 14 cross-sections (EU member states) and 7 cross-section observations - a total of 98 observations per model (inclusive / extractive).

Further, we checked the stationarity for the data used by "Summary" window which provides a detailed view of the results of five stationarity tests and we concluded that is essential to use a method close to the autoregressive distributed lag since the variables proved to be stationary at level and first difference:

✓ Assuming common unit root process (null hypothesis: unit root / alternative: no unit root):

- Levin, Lin & Chu t* (applied in the following assumptions: trend and constant, constant, absence of trend and constant) - 3 results - some disadvantages of the test are: (a) if the number of observations per cross-section is small, the power of the test may be questionable; (b) this test ignores the possibility of the cross-section dependence.

- Breitung t-stat - (applied if the test equation include the trend and constant) - 1 result - this test differs from Levin, Lin & Chu t* since only the autoregressive portion (and not the exogenous components) is removed when constructing the standardized proxies.

✓ Assuming individual unit root process (null hypothesis: unit root / alternative: some cross-sections without unit root):

- Im, Pesaran and Shin W-stat (applied in the following assumptions: trend and constant, constant) - 2 results - this test works better with low number of observations per cross-section than Breitung and have little power when trend is included in the analysis;

- ADF - Fisher Chi-square (applied in the following assumptions: trend and constant, constant, absence of trend and constant) - 3 results - this test allow each cross-section to have different lag length;

- PP - Fisher Chi-square (applied in the following assumptions: trend and constant, constant, absence of trend and constant) - 3 results.

However, the reduced number of observations per cross-section made it necessary to set the maximum lag limit to 1 year and to select the gap between cause and effect according to the empirical evidence.

Also, in case of panel analyses, it is necessary to identify the optimal method of capturing the effects within the model: random effects model or fixed effects model. In this context, we used the Redundant Fixed Effects - Likelihood Ratio test which indicated the use of a fixed-effect model that takes into account the cross-section heterogeneity. Considering the above-mentioned aspects, we started by estimating the following equation:

$$gini_{it} = \alpha_0 + \alpha_1 gini(-1)_{it} + \alpha_2 poverty_{it} + \alpha_3 neetsrate(-1)_{it} + \alpha_4 social_{it} + \alpha_5 creditb_{it} + \theta_{it} + \varepsilon_t \quad (1)$$

where:
- ✓    $i$ = number of countries and $t$ = time period;
- ✓    $\varepsilon_t$ = error term = $\rho_i$ (constant across individuals) + $\theta_{it}$ (composite error term);
- ✓    $\alpha_0$ = the coefficient of the intercept and $\alpha_{0-5}$ = the coefficients of the exogenous variables;
- ✓    $gini$ = Gini coefficient (on a scale of 1 to 100) - Eurostat;
- ✓    $poverty$ = people at risk of poverty rate after social transfers (the share of population earning leass than 60% of the median equivalised national income) - Eurostat;
- ✓    $neetsrate$ = rate of young people neither in employment nor in education and training (the share of population aged 15-24 years neither in employment, or did not received any education or training in the four weeks preceding the survey - Eurostat;
- ✓    $social$ = social expenditures of general government (% of GDP) - Eurostat;
- ✓    $creditb$ = the share of private sector credit of banking sector in GDP - World Bank.

$gini = 1 - \sum_{i=1}^{n}[$*the percentage of the population income of national equivalent income +(the share of population in total population + 2 \* the share of population that is richer than the population under review)$]$*(2)

    In order to apply the fixed effects model, we have added 13 dummy variables in equation (1) these representing the individual intercept for each cross-section (minus one). Therefore, we computed the Least-Squares Dummy Variables (LSDV) estimators for each model as follows:

$$gini_{it} = \beta_0 + \beta_1 gini(-1)_{it} + \beta_2 poverty_{it} + \beta_3 neetsrate(-1)_{it} + \beta_4 social_{it} + \beta_5 creditb_{it} + o_1 dummy_1 + o_2 dummy_2 + o_3 dummy_3 + o_4 dummy_4 + o_5 dummy_5 + \dots o_{13} dummy_{13} + \varepsilon_t \quad (3)$$

where:
- ✓    $\beta_{1-5}$ are the LSDV estimators;
- ✓    $o_{1-13} + \beta_0$ = the intercept for each cross-section;
- ✓    $dummy$ = a binary variable for each cross-section, excepting the last one;
- ✓    the undefined variables was described in equation (1).

Ion Dobre, Ionuț Jianu, Alexandru Bodislav, Carmen Rădulescu, Sorin Burlacu

_______________________________________________________________

However, dummy variables inluced in equation (3) may affect the consistency of the estimators. In this respect, we solved this issue by estimating the following equation, expressing the variables in means terms:

$$\overline{gini}_{it} = \alpha_0 + \alpha_1\overline{gini(-1)}_{it} + \alpha_2\overline{poverty}_{it} + \alpha_3\overline{neetsrate(-1)}_{it} +$$
$$\alpha_4\overline{social}_{it} + \alpha_5\overline{creditb}_{it} + \overline{\theta_{it}} + \overline{\varepsilon_t} \tag{4}$$

Next, we substracted the equation (4) from equation (1):

$$gini_{it} - \overline{gini}_{it} = (\alpha_0 - \alpha_0) + \alpha_1(gini(-1)_{it} - \overline{gini(-1)}_{it}) + \alpha_2(poverty_{it} - \overline{poverty}_{it}) + \alpha_3(neetsrate(-1)_{it} - \overline{neetsrate(-1)}_{it}) + \alpha_4(social_{it} - \overline{social}_{it}) + \alpha_5(creditb_{it} - \overline{creditb}_{it}) + (\theta_{it} - \overline{\theta_{it}}) + (\varepsilon_t - \overline{\varepsilon_t}) \tag{5}$$

orthe final form of the fixed effects model:

$$gini_{it} - \overline{gini}_{it} = \alpha_1(gini(-1)_{it} - \overline{gini(-1)}_{it}) + \alpha_2(poverty_{it} - \overline{poverty_{it}}) + \alpha_3(neetsrate(-1)_{it} - \overline{neetsrate(-1)}_{it}) + \alpha_4(social_{it} - \overline{social_{it}}) + \alpha_5(creditb_{it} - \overline{creditb}_{it}) + \varepsilon_t{}^* \tag{6}$$

where $\varepsilon_t{}^*$ is the pure error from the equation.

Further, we have applied Cross-section weights option and White cross-section covariance estimation method - a method known also as Feasible Generalized Least Squares. Using cross-section weights for both models was intended to allow the existence of heteroscedasticity in low dimensions, if it exists. Other options for ex-ante correction of the heteroskedasticity (Cross-section SUR or Period SUR) were not available due to the small number of observations and the method applied (fixed effects model).

Following the inclusion in the analysis of the 1st-order autoregressive term and of the NEETs rate lagged by 1-year, a number of 84 observations out of 98 possible has resulted. After estimating the models, we have tested the main assumptions needed to validate the maximum verisimilitude of the estimators, as follows: (i) linearity of the model; (ii) model validity - Fisher test; (iii) normal distribution of the residuals - Jarque-Bera test; (iv) absence of heteroskedasticity - Breusch-Pagan-Godfrey test; (v) absence of cross-section dependence - Breusch-Pagan, Pesaran CD and Pesaran scaled LM; (vi) absence of autocorrelation of residuals - Breusch-Pagan test; (vii) absence of multicollinearity - Klein's criterion (verification by calculating the Pearson statistical correlation).

The verification of the absence of autocorrelation of residues and heteroscedasticity was performed by estimating the probability associated with the above-mentioned tests. In order to confirm the absence of heteroskedasticity, we estimated the probability associated with the Breusch-Pagan-Godfrey test, depending on the coefficient of determination of the following equation:

$$residual^2{}_{it} = \gamma_0 + \delta_0 gini(-1)_{it} + \delta_1 poverty_{it} + \delta_2 neetsrate(-1)_{it} + \delta_3 social_{it} + \delta_4 creditb_{it} + \varepsilon_t \quad (7)$$

where $residual^2$ = the square of the standardized residuals from equation (1), $\gamma_0$ is the coefficient of the constant term, $\delta_{0-4}$ are the coefficients of the independent variables and $\varepsilon_t$ is the error term.

Then, we computed the probability of the Breusch-Pagan-Godfrey test using *chisq.dist.rt* function (excel), based on the product of the number of observations (n) of equation (7) and the R-squared, respectively the degrees of freedom (df) which correspond to the number of independent variables except for the constant.

Next, in order to verify the absence of autocorrelation of residuals, we estimated the following equation:

$$residual_{it} = \lambda_0 + \mu_0 gini(-1)_{it} + \mu_1 poverty_{it} + \mu_2 neetsrate(-1)_{it} + \mu_3 social_{it} + \mu_4 creditb_{it} + \mu_5 residual(-1)_{it} + \varepsilon_t \quad (8)$$

where $residual$ = the standardized residuals from equation (1), $residual(-1)$ represents the residuals lagged by 1 year, $\lambda_0$ is the coefficient of the constant term, $\mu_{0-5}$ are the coefficients of the exogenous variables and $\varepsilon_t$ is the error term. Finally, in order to calculate the probability of the Breusch-Pagan test, we followed the same methodology as described above, excepting that in this case we set the number of degrees of freedom to be equal with the number of lags used for the residuals.

## 4. Results and interpretation

In this section we analysed the research results, focusing on identifying the relationship between the quality of institutions and the level of income inequality. In this context, we divided the European Union in two groups of countries, depending on the score of the institutional pillar, published by the World Economic Forum: the cluster of member states with inclusive institutions / extractive institutions. *Figure 1* highlights the institutional specificities of the member states and demonstrates that the structure of the clusters is in line with their level of development. The inclusive institutions cluster is composed by the Northern and Anglo-saxon countries.

The exceptions are EE, MT and PT, whom even if are a part of the catching-up and the Southern submodel of development, these are also categorised in the cluster of member states with inclusive institutions. Regarding the group of countries with extractive institutions, it is composed by the catching-up and the Southern submodel, excepting the cases above-mentioned. Moreover, it can be clearly seen that the structure of the institutional clusters depends on the geographic location, since the Southern Europe (excepting PT) and Eastern Europe constitute the extractive institutions cluster and the Northern Europe (excepting LV and LT) and Western Europe form the inclusive institutions cluster. As regards the relationship between the evolution of the institutions score and the dynamic of the Gini coefficient in the EU-28 member states we computed the Pearson statistical correlation.The result show that there is a negative correlation between their evolution (-37.63%), which determines, preliminarily, the possibility of an inverse relationship between the quality of institutions and Gini coefficient. Following the use of the sub-indices of the institutions pillar as a clustering criterion, both institutional clusters retained a large part of their structure (*Figure 2*).

**Figure 1. The main institutional cluster of EU member states**

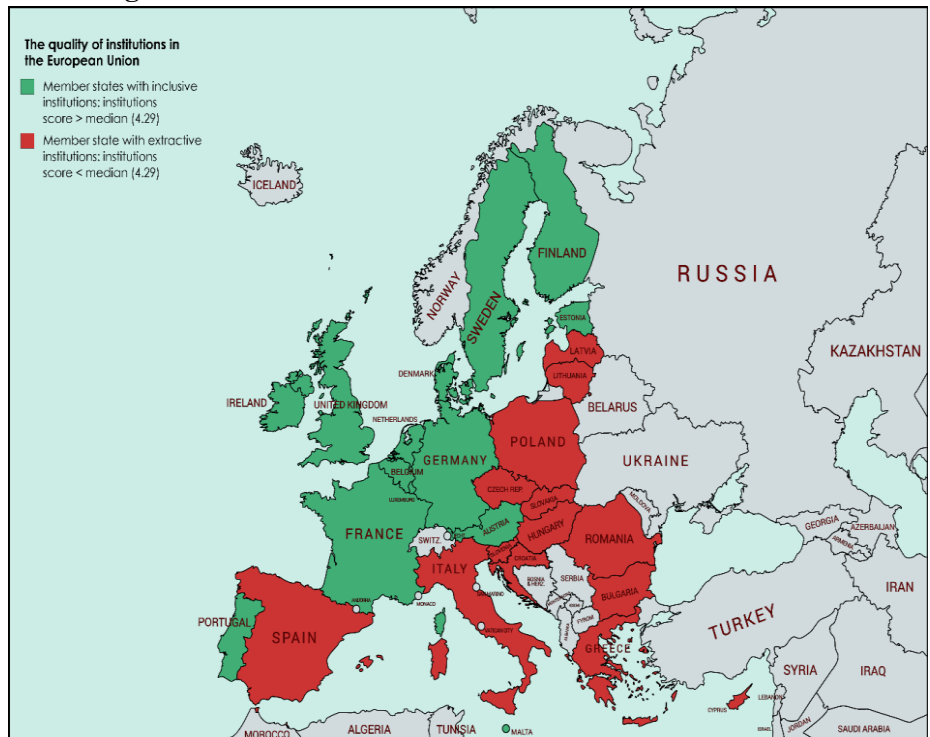

*Source: Own processings on mapchart.net, using World Economic Forum data, The Global Competitiveness Report 2017-2018*

However, when we used the institutions sub-indices as a clustering criterion, few countries remained in the same cluster as the main one: AT, EL, PL, RO, SE. In addition, FI and EE are the countries that are a part of the cluster of member states with inclusive institutions for all sub-indices reviewed, excepting the criteria related to strength of investor protection. At the opposite end were IT and HU, those being categorised in the group of states with inclusive institutions only when using the clustering criterion based on the sub-indices related to the strength of investor protection and the business costs of crime and violence. Other countries have a good institutional position, but have weak positions in terms of the sub-indices associated with security challenges (FR and DE): business costs of crime and violence, organized crime.

The structure of the clusters designed by using the sub-indices as a clustering criteria has proved to be identical with the structure of the main clusters only in the case of four sub-pillars: property rights, public trust in politicians, diversion of public funds, ethical behavior of firms.

**Figure 2. Institutional clustering of EU member states in detail**

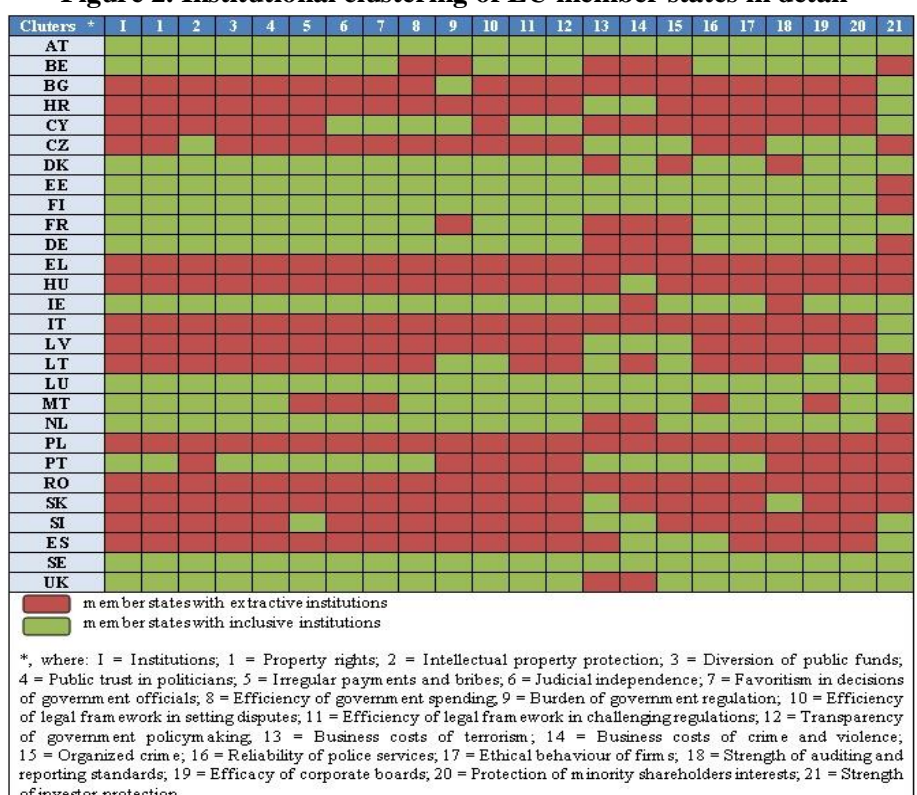

*, where: I = Institutions; 1 = Property rights; 2 = Intellectual property protection; 3 = Diversion of public funds; 4 = Public trust in politicians; 5 = Irregular payments and bribes; 6 = Judicial independence; 7 = Favoritism in decisions of government officials; 8 = Efficiency of government spending; 9 = Burden of government regulation; 10 = Efficiency of legal framework in setting disputes; 11 = Efficiency of legal framework in challenging regulations; 12 = Transparency of government policymaking; 13 = Business costs of terrorism; 14 = Business costs of crime and violence; 15 = Organized crime; 16 = Reliability of police services; 17 = Ethical behaviour of firms; 18 = Strength of auditing and reporting standards; 19 = Efficacy of corporate boards; 20 = Protection of minority shareholders interests; 21 = Strength of investor protection.

*Source: Own processings using World Economic Forum data, The Global Competitiveness Report 2017-2018*

Next, we estimated the impact of the income inequality drivers on the Gini coefficient for each group of countries, which facilitates the comparison of impact differences between them. In order to select the most appropriate estimation method, firstly, we checked the stationarity of the data.

*Table 2* indicates the stationarity of the data at level and at the first difference for both clusters. The optimal lag has been chosen by using Schwarz information criterion, while the final result of the test for each variable has been decided depending on the share of the number of tests confirming stationarity in the total number of tests (12). Only three variables proved to be stationary at level: creditb *, gini (-1) ** and social **; the rest of the variables becoming stationary following the processing of the first difference. These results indicated the use of a autoregressive distributed lag model, but the lag selection was limited by the small number of observations.

**Table 2. Stationarity of the data**

| Stationarity (Schwarz criterion) | Variable | Number of tests confirming stationarity at I(0) | Number of tests confirming stationarity at I(1) |
|---|---|---|---|
| **Inclusive institutions model*** | gini | 3 of 12 | 10 of 12 |
| | gini(-1) | 3 of 12 | 10 of 12 |
| | poverty | 5 of 12 | 9 of 12 |
| | neetsrate(-1) | 5 of 12 | 9 of 12 |
| | social | 2 of 12 | 9 of 12 |
| | creditb | 7 of 12 | Stationary at I(0) |
| **Extractive institutions model**** | gini | 4 of 12 | 11 of 12 |
| | gini(-1) | 7 of 12 | Stationary at I(0) |
| | poverty | 4 of 12 | 11 of 12 |
| | neetsrate(-1) | 5 of 12 | 6 of 6 - I(2) test is forbidden |
| | social | 7 of 12 | Stationary at I(0) |
| | creditb | 5 of 12 | 7 of 12 |

*Source: Own calculations using Eviews 9.0*

Further, we applied the EGLS method mentioned in the methodology, which returned the results displayed in Figure 3. This method was strengthen by the use of a fixed-effect model, taking into account the result of the Redundant Fixed Effects -Likelihood Ratio test (0%), which indicated its use (*Table 3*).

**The impact of historical income inequality on the current one.**The estimation shows that an increase in the Gini coefficient lagged by 1 year with 1 point deviation leads to a rise in the current value of the Gini coefficient by 0.143 deviation points in the case of the cluster of member states with inclusive institutions and by 0.153 deviation points for the group of countries with extractive institutions. The higher impact estimated in the case of the extractive institutions cluster shows that the current income inequalities are more vulnerable to the historical levels of it. In this context, income inequality become more difficult to be controlled due to the existence of extractive institutions that facilitate the increase of income inequality. The sign of the coefficient can be explained by the fact that the population uses its past savings to generate additional income in present or by the higher capital yield than the income yield.

**The impact of the poverty rate on income inequality.**According to the results reported by the inclusive institutions model, an increase in the people at risk of poverty rate by 1pp leads to a change of Gini coefficient by 0.481 deviation points, this being lower than the one reported by the group of countries with extractive institutions (0.512 deviation points).

---

**Figure 3. Estimation results**

Inclusive institutions cluster
Dependent Variable: GINI
Method: Panel EGLS (Cross-section weights)
Sample (adjusted): 2011 2016
Periods included: 6
Cross-sections included: 14
Total panel (balanced) observations: 84
Linear estimation after one-step weighting matrix
White cross-section standard errors & covariance (d.f. corrected)

| Variable | Coefficient | Std. Error | t-Statistic | Prob. |
| --- | --- | --- | --- | --- |
| GINI(-1) | 0.143836 | 0.084348 | 1.705261 | 0.0929 |
| POVERTY | 0.481713 | 0.085019 | 5.665938 | 0.0000 |
| NEETSRATE(-1) | 0.188412 | 0.091272 | 2.064296 | 0.0430 |
| SOCIAL | -0.270828 | 0.073815 | -3.669010 | 0.0005 |
| CREDITB | 0.015937 | 0.004233 | 3.765253 | 0.0004 |
| C | 18.95898 | 0.917521 | 20.66326 | 0.0000 |

Effects Specification

Cross-section fixed (dummy variables)

Weighted Statistics

| | | | |
| --- | --- | --- | --- |
| R-squared | 0.984676 | Mean dependent var | 37.76075 |
| Adjusted R-squared | 0.980432 | S.D. dependent var | 16.21984 |
| S.E. of regression | 0.595624 | Sum squared resid | 23.05995 |
| F-statistic | 232.0342 | Durbin-Watson stat | 1.931053 |
| Prob(F-statistic) | 0.000000 | | |

Unweighted Statistics

| | | | |
| --- | --- | --- | --- |
| R-squared | 0.960550 | Mean dependent var | 28.88333 |
| Sum squared resid | 25.05700 | Durbin-Watson stat | 1.916960 |

Extractive institutions cluster
Dependent Variable: GINI
Method: Panel EGLS (Cross-section weights)
Sample (adjusted): 2011 2016
Periods included: 6
Cross-sections included: 14
Total panel (balanced) observations: 84
Linear estimation after one-step weighting matrix
White cross-section standard errors & covariance (d.f. corrected)

| Variable | Coefficient | Std. Error | t-Statistic | Prob. |
| --- | --- | --- | --- | --- |
| GINI(-1) | 0.153449 | 0.039532 | 3.881677 | 0.0002 |
| POVERTY | 0.512410 | 0.040365 | 12.69428 | 0.0000 |
| NEETSRATE(-1) | 0.123490 | 0.025800 | 4.786393 | 0.0000 |
| SOCIAL | -0.166532 | 0.040715 | -4.090173 | 0.0001 |
| CREDITB | 0.005033 | 0.001933 | 2.603497 | 0.0114 |
| C | 17.47861 | 1.695258 | 10.31029 | 0.0000 |

Effects Specification

Cross-section fixed (dummy variables)

Weighted Statistics

| | | | |
| --- | --- | --- | --- |
| R-squared | 0.997701 | Mean dependent var | 73.46890 |
| Adjusted R-squared | 0.997064 | S.D. dependent var | 67.48472 |
| S.E. of regression | 0.799333 | Sum squared resid | 41.53070 |
| F-statistic | 1566.931 | Durbin-Watson stat | 2.094690 |
| Prob(F-statistic) | 0.000000 | | |

Unweighted Statistics

| | | | |
| --- | --- | --- | --- |
| R-squared | 0.966088 | Mean dependent var | 31.23214 |
| Sum squared resid | 47.97646 | Durbin-Watson stat | 1.614655 |

*Source: Own calculations using Eviews9.0*

Even if, theoretically, poverty can be associated with a low level of inequality between household incomes, in practice the impact of the people at risk of poverty rate on Gini is positive, especially in the case of the EU economy. When a larger share of the population are recording income falls, reaching a lower income level than the poverty line and the population of the last quartile / quintile / decile begins to record an increasing trend of the income share in the national equivalent income, the corresponding scores of the first quartile / quintiles / decile reach low levels, which leads to a lower total score, respectively to a higher Gini coefficient. For a better view of the relationship, the examination of the Gini coefficient formula (*methodology - equation 2*) is essential.

Moreover, poverty and income inequality are also linked through the common set of social policy instruments, the relationship with unemployment or economic growth. The reason why the impact of the increase in the poverty rate on the increase in income inequality is higher in the case of countries with extractive institutions is that, in this group of countries, the rates of people at risk of poverty are higher than the ones recorded by the member states with inclusive institutions.

**The impact of the rate of young people neither in employment nor in education or training on income inequality.** The results have shown a positive relationship between the NEETs rate and the Gini coefficient in both models. An increase in NEETs rate by 1pp rose the level of Gini by 0.188 deviation points after

1 year from the shock, this effect being higher the the one resulted in the case of extractive institutions model (0.123 deviation points).

This indicator includes young unemployment, school dropout, as well as young people who do not participate in vocational training programs or those who have completed a higher education cycle, but do not participate in the labour market. Increasing the unemployment rate leads to a decrease in the income earned by the population and to a rise in the income gaps, given that the income from social benefits is lower than wages.

Moreover, the lack of involvement in vocational training programs and the school dropout limits young people's ability to progress in their careers due to low level of knowledge or insufficient labour market experience. Early school dropout limits, on long-run, the rise of income for early leavers, this category of population being exposed usually to the risk of poverty due to lack of qualifications.

The reason why the impact of the NEETs rate on the Gini coefficient in countries with inclusive institutions is higher consists in the fact that extractive institutions create the premises of a weak relationship between the education system, labour market and career opportunities , which results in a low labour supply elasticity of income. For example, an increase in unemployment will lead also to a similar growth of wages, this effect having the ability to deepen the income inequality. Therefore, the negative correlation in absolute terms between unemployment rate (a component of NEETs rate) and income level is high in countries with inclusive institutions.

**The impact of social spending (expressed as a share of GDP) on income inequality.**The estimation shows that some social policy instruments (such as social spending) are less effective in countries with extractive institutions. In the case of the cluster of member states with inclusive institutions, we have demonstrated that an increase in social spending by 1pp of GDP leads to a drop in the Gini coefficient by 0,270 deviation points, a negative impact significantly higher than the one estimated for the group of countries with extractive institutions 0.166 deviation points).

Social spending is more effective in countries with inclusive institutions as they are conditioning social benefits on labour market participation. Other countries provide some financial incentives to reinforce the young people's motivation to participate and perform in school / student activities. However, some measures are losing their impact due to the structural challenges faced by the educational systems from the countries with extractive institutions. Moreover, regarding this indicator, we have also found a mentality disparity. In the Northern, Continental and Anglo-saxon sub-models, the population prefers to use unemployment benefits to identify new career opportunities for starting their activity on the labour market, while the population from the member states with extractive institutions are losing their motivation for labour market participation, some of them being content with these benefits. Paradoxically, although the Nordic countries provides high unemployment benefits, the population is oriented towards a participatory life, which leads to a high employment rates in these countries - a good example is SE with the highest employment rate for the 20-64 age-group in the EU-28 (81.8% in 2017).

**The impact of the private sector credit granted by the banking sector on income inequality.**Regarding the implications of banking activity on income inequality, we found that the increase of private sector credit granted by banks by 1pp of GDP leads to the growth of Gini coefficient by 0.015 deviation points in member states with inclusive institutions, which is higher than the impact associated to the extractive institutions model (0.005 deviation points). The sign of the coefficient can be explained by the higher accessibility of the individuals / corporates earning high incomes / profits, these type of loans being less risky for banks.

If the beneficiary is an economic entity, it may access funds in order to establish new development strategies, which ultimately results in an increase in employee profits and earnings that facilitates an increase in the income gap between employees of that company and those of other companies that have less access to credit. The reason why the impact of private sector credit proved to be more robust in the case of the group of countries with inclusive institutions consists in the fact that, in these countries, the conditions for accessing the loans are tighter than those in force in countries with extractive institutions. Therefore, big economic players or the high income population have a greater accessibility to loans in the countries with inclusive institutions.

**Testing hypotheses.**As can be seen in *Figure 3*, the evolution of the regressors explains 98.46% of the dynamic of the Gini coefficient for the cluster of states with inclusive institutions. Even if, the coefficient of determination is higher in the extractive institutions model (99.77%), the differences are slight. This made it possible the validation of the proper selection of the regressors in both models. All coefficients are statistically significant excepting the coefficient of the autoregressive term associated to the inclusive institutions model that is statistically significant only at 10% significance threshold. Moreover, the probabilities associated with the Fisher test ($<5\%$) confirmed the validity of both models.

All tests performed checking the absence of cross-section dependence (excepting Breusch-Pagan LM test - the inclusive institutions model - *Table 3*) confirmed the null hypothesis. Since most tests have confirmed this assumption, we have accepted the hypothesis according to which there is no dependence between cross-sections.

The distribution of the residuals proved to be normal since the probability associated to Jarque-Bera statistics is greater than 5% (21.58% - the inclusive institutions model, 43.25% - the extractive institutions model). Following the estimation of Equation 3, the results of the Breusch-Pagan-Godfrey test confirmed the homoskedastic character (the constant variance) of the residuals for both models, as the probabilities were above the 5% significance threshold.

The probabilities associated to Breusch-Pagan serial correlation test led to the acceptance of the null hypothesis (there is no autocorrelation between the residuals) in both models.

_______________________________________________________________

### Tabel 3. Assumptions checked

| Assumptions checked | Test | Inclusive institutions model (probability) | Extractive institutions model (probability) |
|---|---|---|---|
| Fixed effects model is redundant | Redundant Fixed Effects - Likelihood Ratio test | 00.00% | 00.00% |
| Random effects model is appropriate | Correlated Random Effects - Hausman test | 100.00% | 100.00% |
| Absence of cross-section dependence | Breusch-Pagan LM | 03.51% | 07.08% |
| | Pesaran scaled LM | 37.83% | 61.75% |
| | Bias-corrected scaled LM | 60.38% | 36.04% |
| | Pesaran CD | 87.91% | 71.85% |
| Normal distribution of the residuals | Jarque-Bera test | 21.58% | 43.25% |
| Absence of serial correlation | Breusch-Pagan serial correlation test | 24.10% (n = 70, df = 1) | 24.88% (n = 70, df = 1) |
| Homoskedasticity | Breusch-Pagan-Godfrey Heteroskedasticity test | 75.24% (n = 84, df = 5) | 97.71% (n = 84, df = 5) |

*Source: Own calculations using Eviews 9.0*

The validated assumptions confirmed the maximum verisimilitude of the estimators. Therefore, we accepted the robustness of the coefficients.

## 5. Conclusions and recommendations

The designed clusters are in line with the EU's sub-models of development, which highlights the link between the quality of institutions and the level of economic development. Based on the analysis of the statistical correlation between income inequality and the institutions pillar, we identified a negative correlation of approximately 37%, which reflects a possible negative effect of institutions on income inequality.

This paper showed that inequality is more persistent in countries with extractive institutions, given the higher impact of historical inequality on the current one than the impact estimated for the extractive institutions model. Also, the Gini coefficient reacts higher to a change in the dynamics of the people at risk of poverty rate in countries with low institutions quality. Moreover, the extractive institutions model is weaker than the inclusive model as regards the efficiency of government intervention through social protection expenditure. Our estimates proved that the social spending is more efficient in European Union member states having a favorable institutional climate.

The impact of private sector credit granted by the banking sector on the Gini coefficient dynamics has proved to be positive in both models. Our estimations demonstrated that the effect of private sector lending on inequality is higher in the inclusive institutions model than the one associated to the extractive institutions model as a consequence of the tighter lending conditions in the first mentioned group.

Our paper confirmed the positive relationship between the rate of young people neither in employment nor in education or training (NEETs) and income inequality in both models. The results of the estimation indicated a higher impact of the NEETs rate on inequality in countries with inclusive institutions, which can be explained by the stronger link between wage levels and labour supply in these countries.

The issues encountered in countries with extractive institutions can be solved by improving government efficiency and the quality of institutions as well as by strengthening the role of trade unions or social partners in order to increase the bargaining power of them. Moreover, acts of corruption affect all sub-indices of the institutional pillar and its decrease may have favorable consequences on reducing income inequality.Identifying mechanisms to monitor and reduce the influence of lobby groups in the political decision-making process to obtain benefits and preferential treatment is a good way forward.Finally, we have identified an inclusive growth dashboard covering threedimensions on which the national governments should focus on in order to streamline the impact of government intervention on moderating income inequality:

- participatory life: social benefits conditioned by labour market participation or educational outcomes, tight rules for corporates in order to improve the behaviour of firms, public-private partnerships between universities and economic entities, improving health systems and budgeting them depending on their needs, encouraging active ageing, increasing the labour market flexibility, promoting career opportunities for young people, strengthening the role of trade unions;
- governance: improving property rights, intensifying the punishment for corruption, improving the efficiency of government spending, reducing the burden of government regulation, increasing the incidence of progressive tax, improving transparency, strengthening auditing and reporting standards, labour market institutions (minimum wage setting policies / employment security policies / employment protection legislation), anti-discrimination institutions;
- money: low inflation, new financial instruments for small and medium enterprises or low income population, progressive interest rates depending on the income earned, adjustment funds for absorbing digital shocks;